\documentclass[10pt,aps,twocolumn,showpacs]{revtex4}
\usepackage{graphicx}

\begin{document}
\title{Magnetotransport in dilute 2D Si-MOSFET system}

\author{M. V. Cheremisin}
\affiliation{A.F.Ioffe Physical-Technical Institute,
St.Petersburg, Russia}
\date{\today}

\begin{abstract}
The beating pattern of Shubnikov-de Haas oscillations is
reproduced in both the crossed and tilted magnetic field
configuration and in presence of zero-field valley splitting in
Si-MOSFET system. The consequences of IQHE in extremely dilute
2DEG are discussed.
\end{abstract}


\maketitle

A great deal of interest has been focussed on the anomalous
transport behavior\cite{Abrahams} of a wide variety of low-density
2D systems. It has been found that, below some critical density,
the cooling causes an increase in resistivity, whereas in the
opposite high density case the resistivity decreases. Another
unusual property of low-density 2D systems is their response to
perpendicular magnetic field. In dilute Si-MOSFET system the spin
susceptibility known to be strongly enhanced, therefore results in
magnetotransport features associated mostly with spin. Although
numerous theories have been put forward to account for these
effects, the origin of the above behavior is still the subject of
a heated debate. In present paper we investigate the beating
pattern of Shubnikov-de Haas(SdH) oscillations caused by
zero-field valley splitting in Si-MOSFET system. Then, we analyze
the SdH beating pattern for the crossed magnetic field
configuration case. Magnetotransport in extremely dilute 2DEG
subjected into quantizing magnetic fields is discussed.

In contrast to conventional SdH formalism extensively used to
reproduce low-field data we allude to alternative
approach\cite{Cheremisin3} seems to give an overwhelming efforts
to resolve magnetotransport problem within both SdH and IQHE
regimes. Based on a thermodynamic approach, in
Ref.\cite{Cheremisin3} has been calculated the magnetoresistivity
of a 2D electron gas, assumed nevertheless dissipationless in a
strong quantum limit. Standard measurements, with extra current
leads, define the magnetoresistivity caused by a combination of
Peltier and Seebeck effects.\cite{Kirby},\cite{Cheremisin} The
current causes heating(cooling) at the first(second) sample
contacts, due to the Peltier effect. The contact temperatures are
different. The measured voltage is equal to the Peltier
effect-induced thermoemf which is linear in current. As a result,
the magnetoresistivity is non-zero as $I\rightarrow 0$. The
resistivity found to be a universal function of magnetic field and
temperature, expressed in fundamental units $h/e^{2}$.

The Si-MOSFET energy spectrum modified with respect to valley and
spin splitting yields
\begin{equation}
\varepsilon_{n}=\hbar\omega_{c}(n+1/2)\pm \frac{\Delta_{s}}{2} \pm
\frac{\Delta_{v}}{2} \label{spectrum}
\end{equation}
where $n=0,1..$ is the LL number, $\omega_{c}=
\frac{eB_\perp}{mc}$ the cyclotron frequency,
$\Delta_{s}=g^{*}\mu_{B}B$ the Zeeman splitting, $g^{*}$ the
effective g-factor, $B=\sqrt{B^{2}_{\perp}+B^{2}_{\parallel}}$ the
total magnetic field. Then,
$\Delta_{v}$[K]$=\Delta^{0}_{v}+0.6B_{\perp}$[T] is the density
independent\cite{Pudalov3} valley splitting. In contrast to valley
splitting, the spin susceptibility $\chi=\frac{g^{*}m}{2m_{0}}$(
here, $m_{0}$ is the $free$ electron mass ) known to exhibit
strong enhancement upon 2D carrier depletion. The latter result is
confirmed independently by magnetotransport measurements in tilted
magnetic field \cite{Fang},\cite{Okamoto}, perpendicular
field\cite{Kravchenko1} and beating pattern of SdH oscillations
\cite{Pudalov5} in crossed fields.

Recall that in strong magnetic fields $\hbar \omega_{c} \gg
kT,\hbar/\tau$ the electrons can be considered dissipationless,
therefore $\sigma_{xx}, \rho_{xx} \simeq 0$. Here, $\tau$ is the
momentum relaxation time. Under current carrying conditions the
only reason for finite longitudinal resistivity seems to be
thermal correction mechanism discussed in Ref.\cite{Kirby}.
Following Ref.\cite{Cheremisin3} one obtains
\begin{equation}
\rho= \rho_{yx}\frac{\alpha^2}{L} \label{magnetoresistivity}
\end{equation}
where $\alpha $ is the thermopower, $\rho
_{yx}^{-1}=Nec/B_{\perp}$ the Hall resistivity, $N=- {\partial
\Omega \overwithdelims()
\partial \mu }_{T}$ the 2D density, $\Omega=-kT \Gamma \sum \limits_{n}\ln
\left(1+\exp \left(\frac{\mu -\varepsilon_{n}}{kT}\right)\right)$
the thermodynamic potential modified with respect to
abovementioned energy spectrum, $\Gamma=\frac{eB_\perp}{hc}$ the
zero-width LL DOS. In actual fact, in strong magnetic fields 2D
thermopower is a universal quantity\cite{Girvin}, proportional to
the entropy per electron: $\alpha =-{\frac{S}{eN}}$, where
$S=-{\partial \Omega \overwithdelims()
\partial T }_{\mu}$ is the entropy. Both $S,N$, and, thus $\alpha,\rho$
are universal functions of $\xi$ and the dimensionless magnetic
field $\hbar \omega _{c}/\mu =4/\nu$, where $\nu=N_{0}/\Gamma$ is
the conventional filling factor, $N_{0}=\frac{2m \mu}{\pi \hbar
^{2}}$ is the zero-field density of strongly degenerate 2DEG.

Using Lifshitz-Kosevich formalism, asymptotic formulae can be
easily derived for $N,S$, and hence for $\rho_{yx},\rho$, valid
within low temperature, magnetic field limit $\nu ^{-1},\xi <1$:
\begin{eqnarray}
N=N_{0}\xi F _{0}(1/\xi )+2\pi \xi
N_{0}\sum\limits_{l=1}^{\infty}\frac{(-1)^{l}\sin (\frac{\pi
l\nu}{2})}{\sinh (r_{l})}R(\nu),
\label{Lifshitz} \\
S=S_{0}-2\pi ^{2}\xi kN_{0}\sum\limits_{l=1}^{\infty }(-1)^{l}\Phi
(r_{l})\cos \left(\frac{\pi l\nu}{2}\right)R(\nu), \nonumber
\end{eqnarray}
where $S_{0}= kN_{0}\frac{d}{d\xi }\left[\xi ^{2}F_{1}(1/\xi)
\right]$ is the entropy at $B_{\perp}=0$, $F _{n}(z)$ the Fermi
integral, $\Phi (z)=\frac{1-z\coth (z)}{z\cdot \sinh (z)}$,
$r_{l}=\pi ^{2}\xi\nu l/2$ the dimensionless parameter. Then,
$R(\nu)=\cos(\pi l s)\cos(\pi lv)$ is the form-factor,
$s=\frac{\Delta_{s}}{\hbar\omega_{c}}=\chi \frac{B}{B_{\perp}}$
the dimensionless Zeeman spin splitting,
$v=\frac{\Delta_{v}}{\hbar\omega_{c}}=\frac{\Delta^{0}_{v}
\nu}{4\mu}+0.12$ the dimensionless valley splitting.

\begin{figure} \vspace*{0.5cm}
\includegraphics[scale=0.75]{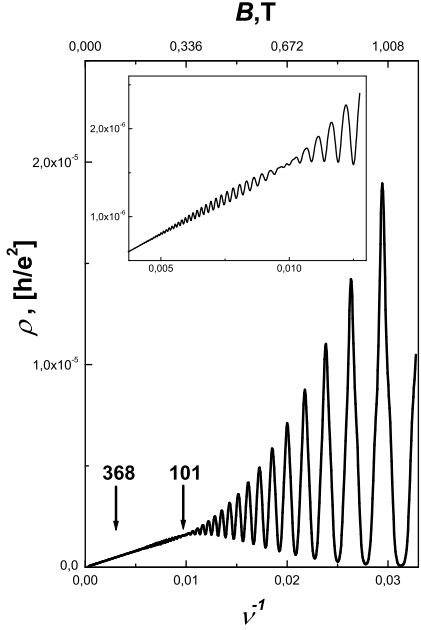}
\caption[]{\label{f.4} SdH oscillations at $T=0.3$K for Si-MOSFET
sample\cite{Pudalov4}: $N_{0}=8.39*10^{11}$ cm$^{-2}$, spin
susceptibility $\chi=0.305$ and valley splitting
$\Delta_{v}$[K]=$\Delta^{0}_{v}+0.6B_{\perp}$[T]. Zero-field
valley splitting $\Delta^{0}_{v}=0.92K$ is a fitting parameter.
Arrows depict the beating nodes at $i=1,3$. Inset: the enlarge
plot of the beating node from the main panel.} \vspace*{-0.5cm}
\end{figure}

Let us first consider zero-$B_{\parallel}$ case, when the Zeeman
spin splitting is reduced to field-independent constant, i.e.
$s=\chi$. Then, in low-$T,B_{\perp}$ limit the valley splitting
$\Delta^{0}_{v}$ known to be resolved\cite{Pudalov4}, therefore
leads to beating of SdH oscillations. For actual first-harmonic
case( i.e. $l=1$ ), the beating nodes can be observed when
$\cos(\pi v)=0$, or $\nu^{v}_{i}=\frac{4\mu
(i/2-0.12)}{\Delta^{0}_{v}}$, where $i=1,3..$ is the beating node
index. For 2DEG parameters( see Fig.\ref{f.4}) reported in
Ref.\cite{Pudalov4} we estimate $\nu^{v}_{1}=101$, therefore
$\Delta^{0}_{v}=0.92$K. The second node is expected to appear at
$\nu^{v}_{2}= 368$. However, SdH oscillations are, in fact,
resolved when $\nu \leq 1/\xi=203$, therefore the second beating
node was not observed in experiment\cite{Pudalov4}. Moreover, the
observed disappearance of the first beating node upon 2D carrier
depletion $N<3*10^{11}$ cm$^{-2}$ is governed by the same
condition because in this case $\nu \leq 1/\xi=73$ being of the
order of the first beating node. Note that suppression the beating
nodes at higher densities($N>9*10^{11}$ cm$^{-2}$) reported in
Ref.\cite{Pudalov4} is, however, unexpected within our simple
scenario.

\begin{figure} \vspace*{0.5cm}
\includegraphics[scale=0.75]{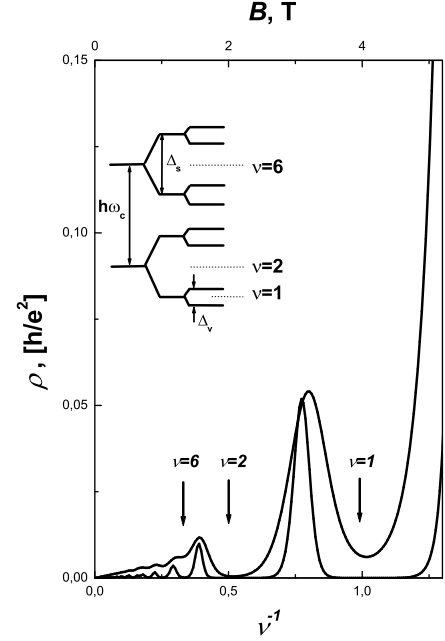}
\caption[]{\label{f.5} Magnetoresistivity at $T=0.36;0.18$K for
dilute 2DEG Si-MOSFET\cite{Kravchenko2}: $N_{0}=10^{11}$
cm$^{-2}$, spin susceptibility $\chi=0.5$ and valley splitting
denoted in Fig.\ref{f.4}. Inset: Energy spectrum specified by
Eq.\ref{spectrum} for two lowest LLs} \vspace*{-0.5cm}
\end{figure}

\begin{figure} \vspace*{0.5cm}
\includegraphics[scale=0.75]{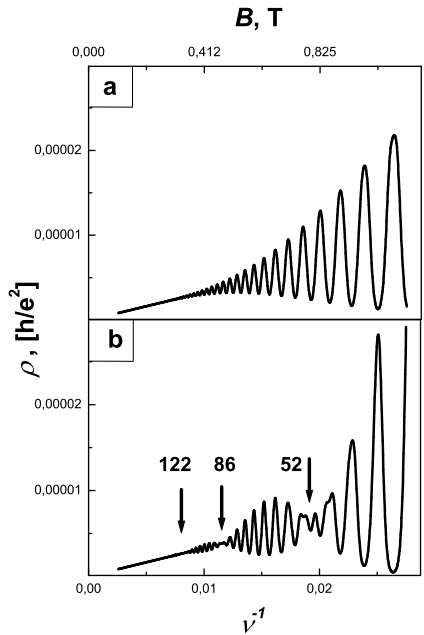}
\caption[]{\label{f.6} SdH beating pattern oscillations at
$T=0.35$K for Si-MOSFET sample\cite{Pudalov5}:
$N_{0}=10.6*10^{11}$ cm$^{-2}$, spin susceptibility $\chi=0.27$,
$\Delta^{0}_{v}=0$ and a)$B_{\parallel}=0$ b)$B_{\parallel}=4.5$T
($\nu_{\parallel}=9.25$). Arrows depict the beating nodes at
$j=3,5,7$} \vspace*{-0.5cm}
\end{figure}

We now analyze the case of low-density 2D system in strong
magnetic field with the only lowest LLs occupied. For extremely
dilute 2DEG($N \simeq 10^{11}$ cm$^{-2}$) the energy spectrum(
Fig.\ref{f.6}, inset) known to be strongly affected by enhanced
spin susceptibility. In contrast to high density case with
cyclotron minima occur at $\nu=4,8,12..$, in dilute 2DEG the only
spin minima( $\nu=2,6,10..$ ) are observed.\cite{Kravchenko2} As
expected, the spin (cyclotron) minima fillings are proportional to
the odd( even ) numbers multiplied by factor of two due to the
valley degeneracy. In stronger fields magnetoresistivity data
exhibit $\nu=1$ minimum associated with valley splitting. With the
help of energy spectrum implied by Eq.(\ref{spectrum}) one can
easily find that the last minima occur when the Fermi level lies
between the lowest valley-split LLs, i.e.
$\mu=\hbar\omega_{c}(1-\chi)/2$. The sequence of minima at
$B=4,2,0.66$T reported in Ref.\cite{Kravchenko2} provides the
independent test for spin susceptibility in high-$B_{\perp}$
limit. In Fig.\ref{f.5} we represent the magnetoresistivity
specified by Eq.\ref{magnetoresistivity} and then use $\chi=0.5$
in order to fit the observed minima sequence. Surprisingly, the
value of spin susceptibility is lower than that $\chi=0.86$
extracted from crossed-field SdH beating pattern
analysis\cite{Pudalov5}. We attribute the above discrepancy to,
for example, the possible magnetic field dependence of spin
susceptibility.

Finally, we focus on magnetotransport problem in crossed magnetic
field configuration. Following experiments\cite{Pudalov4} we
further neglect the zero-field valley splitting for actual high
density case($N>9*10^{11}$ cm$^{-2}$). At fixed parallel magnetic
field the dimensionless Zeeman splitting yields $s=\chi
\sqrt{1+\nu^{2}/\nu^{2}_{\parallel}}$, where we introduce an
auxiliary "filling factor" $\nu_{\parallel}=\frac{hcN_{0}}{
eB_{\parallel}}$ associated with the parallel field. Within
low-$B_{\perp}$ limit the parallel field induced spin splitting
result in the beating of SdH oscillations as well. One can easily
derive the condition for SdH beating nodes as follows $\cos(\pi
s)=0$ or $\nu_{j}^{s}=\nu_{\parallel}\sqrt{(j/2 \chi)^2-1}$, where
$j=1,3...$. The sequence of the beating nodes observed in
Ref.\cite{Pudalov5} allowed the authors to deduce the density
dependence of the spin susceptibility. As an example, for 2DEG
parameters\cite{Pudalov5} in Fig.\ref{f.6} we reproduce the
magnetoresistivity implied by
Eqs.(\ref{magnetoresistivity}),(\ref{Lifshitz}). The phase of SdH
oscillations remains the same between the adjacent beating nodes,
and changes by $\pi$ through the node in consistent with
experiments.

\begin{figure} \vspace*{0.5cm}
\includegraphics[scale=0.75]{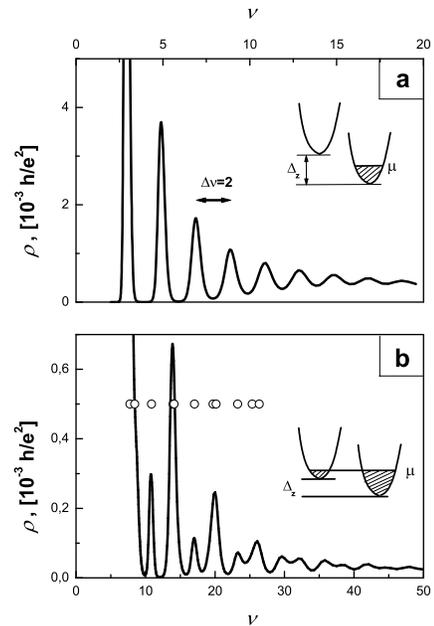}
\caption[]{\label{f.7} Small-angle SdH oscillations at $T=1.35$K
for Si-MOSFET system\cite{Vitkalov}: a) spin polarized electrons(
$p=1.01$) at $N_{0}=3.72*10^{11}$ cm$^{-2}$, spin susceptibility
$\chi=0.42$\cite{Pudalov5}, "effective filling factor"
$\nu_{tot}=0.83$ and b)partially polarized case($p=0.29$) at
$N_{0}=9.28*10^{11}$ cm$^{-2}$, spin susceptibility
$\chi=0.30$\cite{Pudalov5} and $\nu_{tot}=2.06$. Maxima positions
are represented by open dots. Insets: schematic band diagrams at
$B=B_{\parallel}$} \vspace*{-0.5cm}
\end{figure}

We now consider 2DEG magnetotransport in tilted configuration with
the sample rotated in a constant magnetic field
\cite{Fang,Okamoto,Vitkalov}. In this case, the SdH beating
pattern known to depend on the spin polarization degree
$p=\frac{\Delta_{s}}{2 \mu}=\frac{2\chi}{\nu_{tot}}$, where we
introduce the auxiliary "filling factor"
$\nu_{tot}=\frac{hcN_{0}}{eB}$, associated with the total magnetic
field. Conventionally, the spin polarization degree is related to
parallel field $B_{c}$ required for complete spin polarization,
therefore $p=\frac{B}{B_{c}}$. Performing a minor modification in
Eq.\ref{magnetoresistivity}, namely that $s=\chi
\frac{\nu}{\nu_{tot}}$, in Fig.\ref{f.7} we represent the
magnetoresistivity as a function of filling factor for 2DEG plane
rotated with respect to constant magnetic field B=18T (see
Ref.\cite{Vitkalov}). For simplicity, we omit zero-field valley
splitting. Then, arguing the LLs spreading is neglected within our
simple approach, we use somewhat higher temperature compared to
that in experiment\cite{Vitkalov}. For spin polarized system SdH
oscillations($p=1.01$ Fig.\ref{f.7}) is caused by the only lowest
valley-degenerated spin-up subband. At low temperatures, the
valley-splitting associated deep at $\nu=3$ found to be resolved.
With the help of energy spectrum, specified by Eq.\ref{spectrum},
the high-filling maxima occur at $\frac{4(N+1/2)}{1+p}\sim 2N+1$,
therefore have a period $\Delta \nu=2$. In contrast, the partially
polarized high-density 2DEG case($p=0.29$) depicted in
Fig.\ref{f.7},b demonstrates rather complicated beating pattern
caused by the both spin-up and spin-down subbands. One can easily
demonstrate that high-filling maxima occur at $\frac{4(N+1/2)}{1
\pm p}$ (dots in Fig.\ref{f.7},b ), thus depend on spin
polarization degree. The ratio of oscillation frequencies of two
spin subbands is $f^{\downarrow}/f^{\uparrow}=\frac{1-p}{1+p}$
being consistent with experiment \cite{Vitkalov}. At a moment, we,
however, cannot explain the puzzling behavior of low-filling
magnetoresistivity known(Ref.\cite{Okamoto},\cite{Kravchenko2}) to
be insensitive to parallel field component.

We emphasize that the data represented in Fig.\ref{f.4}-\ref{f.7}
differs with respect to those provided by conventional formalism
in the following aspects: i) low-field( $\omega_{c}\tau \leq 1 $)
quantum interference and classical negative magnetoresistivity
background is excluded within our approach and ii) in contrast to
conventional SdH analysis, our approach determines(at
$\omega_{c}\tau \gg 1 $) the $absolute$ value of
magnetoresistivity, and, moreover provides the continuous
transition SdH-to-QHE regime($\hbar\omega_{c} \gg kT$). Minor
point is that our approach predicts somewhat lower SdH
oscillations amplitude compared to that in experiment. However, in
IQHE regime the magnitude of magnetoresistivity is well
comparable( see Ref.\cite{Cheremisin3}) with experimental values.

In conclusion, we demonstrate the relevance of the approach
suggested in Ref.\cite{Cheremisin3}) regarding to low-field
beating pattern SdH oscillations in both crossed and tilted
magnetic field configuration. Then, we examine the features
concerning IQHE in dilute Si-MOSFET system.

This work was supported by RFBR(grant 03-02-17588), and
LSF(HPRI-CT-2001-00114, Weizmann Institute)

\end{document}